\documentclass[aps,prc]{revtex4}
\usepackage{epsfig}
\newcommand{\vsig}{\mbox{\boldmath$\sigma$\unboldmath}}
\newcommand{\veps}{\mbox{\boldmath$\epsilon$\unboldmath}}

\newcommand{\be}{\begin{equation}}
\newcommand{\ee}{\end{equation}}
\newcommand{\bea}{\begin{eqnarray}}
\newcommand{\eea}{\end{eqnarray}}
\newcommand{\bean}{\begin{eqnarray*}}
\newcommand{\eean}{\end{eqnarray*}}

\newcommand{\gapproxeq}{\lower
.7ex\hbox{$\;\stackrel{\textstyle >}{\sim}\;$}}
\newcommand{\lapproxeq}{\lower
.7ex\hbox{$\;\stackrel{\textstyle <}{\sim}\;$}}

\begin{document}

\title{\bf Double polarization asymmetry as a possible filter for 
$\Theta^+$'s parity}

\author{Qiang Zhao\footnote{e-mail: Qiang.Zhao@surrey.ac.uk}
and J.S. Al-Khalili
}
\affiliation{ Department of Physics, 
University of Surrey, Guildford, GU2 7XH, United Kingdom}

\begin{abstract}
We present an analysis of the Beam-Target double polarization asymmetry
for the photoproduction of $\Theta^+$ in $\gamma n\to \Theta^+ K^-$.
We show that this quantity can serve as a filter for the determination
of the $\Theta^+$'s spin-parity assignment near threshold.  It is
highly selective between $1/2^+$ and $1/2^-$ configurations due to
dynamical reasons.

\end{abstract}

\maketitle  
\vskip 1.cm


%\section{Introduction}

The newly-discovered exotic $\Theta^+$, with strangeness $S=+1$, has
stimulated a great number of theoretical discussions during the past
few months. Although more experimental data are strongly recommended to
confirm its properties, the significance of this discovery is that, for
the first time, the existence of unconventional baryons is more than a
purely theoretical prediction.  So far, several experimental groups
have seen this narrow state
independently~\cite{spring-8,diana,clas,saphir,clas-2}, while another
state ($\Xi^{--}$) was also reported by NA49 Collaboration
recently~\cite{na49}.

Concerning the nature of the $\Theta^+$, its rather-low mass (1.535
GeV) and narrow width ($<25$ MeV) makes it an attractive candidate for
the SU(3) Skyrme model predicted $\bar{\bf 10}$
multiplets~\cite{skyrme,dpp}.  On the other hand, the existence of such
an $S=+1$ baryon does not rule out the conventional quark model, where
baryons can be classified perfectly into ${\bf 8}$, ${\bf 10}$ and a
singlet ${\bf 1}$.  Therefore, if an extra quark pair is present, e.g.
a $u\bar{s}$ within a neutron ($udd$) or a $d\bar{s}$ within a proton
($uud$), a $\Theta^+$ of pentaquark $uudd\bar{s}$ can be constructed.
However, such a treatment implies the extension of the SU(3) flavor
space. For instance, multiplets of anti-decuplet $\bar{\bf 10}$,
27-plet {\bf 27}, and 35-plet {\bf 35}, are possible, and will lead
to different predictions in comparison with the Skyrme model.

The spin and parity of the exotic $\Theta^+$ baryon turns out to be an
essential issue for understanding the underlying dynamics. A series
of theoretical studies exploring the underlying dynamics have been
carried out, for instance, in the Skyrme
model~\cite{bfk,prasza-1,praszalowicz,cohen,ikor,cohen-2,wu-ma}, quark
models~\cite{gao-ma,JW,k-l,s-r,caps,carlson,cheung,glozman,jm,hzyz,okl,dp,bijker,
l-h-d-c-z},
and other phenomenologies~\cite{kk,gk}, which rely crucially
on the phenomenological assumptions for the $\Theta^+$'s quantum
numbers.  QCD sum rule studies~\cite{zhu,mnnrl,sdo,h-d-c-z}, and
lattice QCD calculations~\cite{sasaki,cfkk} are also reported.  This
situation suggests explicit experimental
confirmation of the quantum numbers of the $\Theta^+$ is not only
important for establishing its status on a fundamental
basis~\cite{kishimoto,juengst}, but also important for any progress in
understanding its nature, and the existence of its other
partners~\cite{close1,d-c}.  

A number of theoretical studies of the
$\Theta^+$ in meson photoproduction and meson-nucleon scattering were
made recently concentrating on cross section
predictions~\cite{pr,hyodo,nam,liu-ko,oh,yu}.  However, due to the lack
of knowledge about the underlying reaction mechanism, such
studies of the reaction cross sections are strongly
model-dependent.  For instance, the total width of $\Theta^+$ still has
large uncertainties and could be much
narrower~\cite{asw,aasw,nussinov,gothe,hk,cckn}, 
and the role played by
$K^*$ exchange, as well as other {\it s}- and {\it u}-channel processes
are unknown. Also, in a phenomenological approach the energy dependence of the
couplings is generally introduced into the model via empirical form
factors, which will bring further uncertainties. Taking this into account, 
there are advantages with
polarization observables (e.g.  ambiguities arising from the unknown
form factors can be partially avoided). In association with the cross
section studies, supplementary information about the $\Theta^+$ can be
obtained~\cite{zhao-theta,nakayama,okl-2}.  We also note that an
interesting method to determine the $\Theta^+$'s parity was recently
proposed in Ref.~\cite{thh}, and further detailed in
Ref.~\cite{hanhart}.

In this letter, we will show that one of the beam-target (BT)
double-polarization-asymmetry observables will be more selective 
to the $\Theta^+$'s parity. It is very likely that 
the BT asymmetry would serve as a parity filter for the $\Theta^+$
near its production threshold. We will concentrate on the spin-parities of
the $\Theta^+$ of $1/2^+$ and $1/2^-$, while some discussions
will be devoted to the possible $3/2$ partner.

%%%%-- Parity positive

\begin{figure}
\begin{center}
\epsfig{file=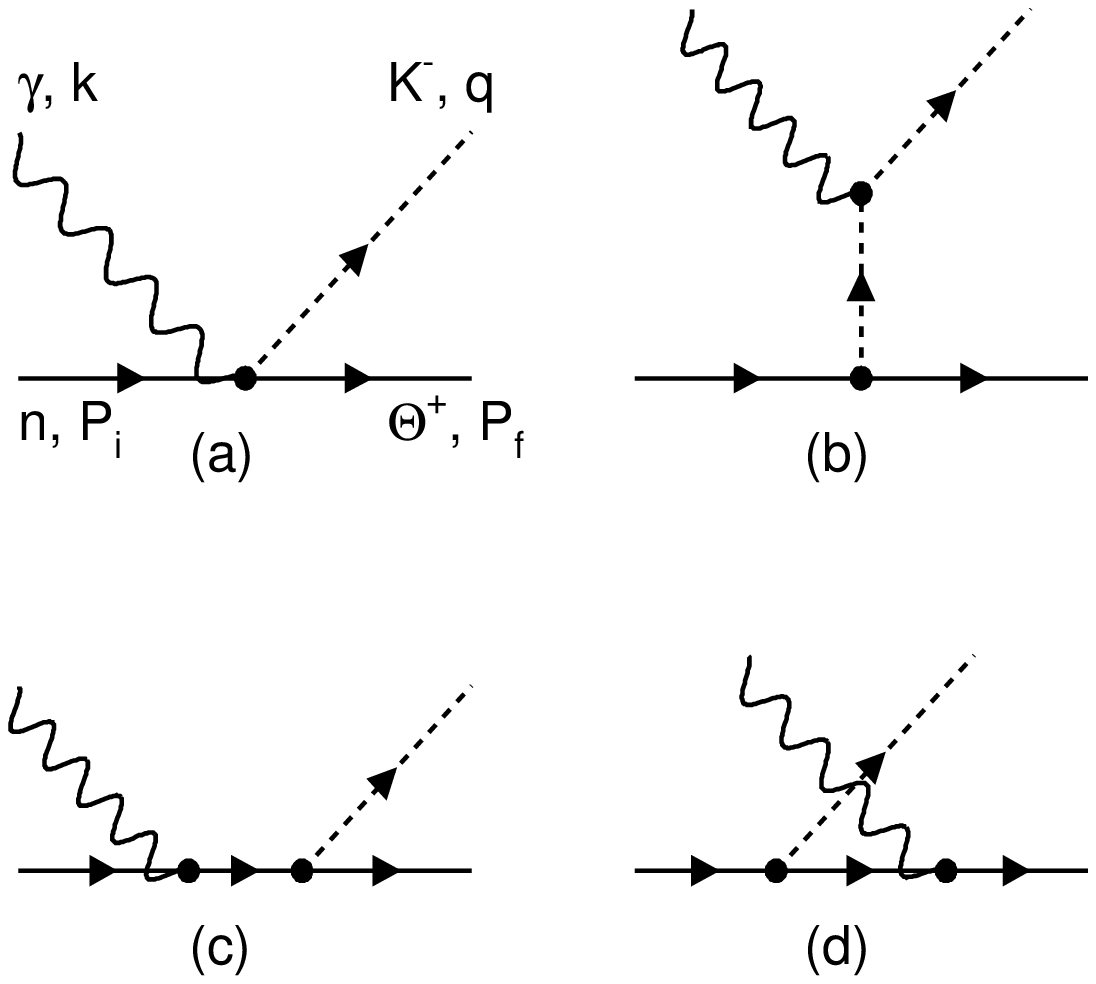, width=10cm,height=9.cm}
\caption{Feynman diagrams for $\Theta^+$ photoproduction in the Born 
approximation. 
}
\protect\label{fig:(1)}
\end{center}
\end{figure}

For a positive parity $\Theta^+$ ($1/2^+$), a pseudovector effective Lagrangian 
is introduced for the $\Theta NK$ coupling~\cite{zhao-theta}.
In $\gamma n\to K^-\Theta^+$, 
four transition amplitudes labelled by the Mandelstam variables
will contribute in the Born approximation limit
as shown by diagrams in Fig.~\ref{fig:(1)}: the contact term, {\it t}-channel 
kaon exchange, {\it s}-channel nucleon exchange and {\it u}-channel $\Theta^+$
exchange.
The transition can be expressed as:
\be
{\cal M}_{fi}=M^c + M^t + M^s + M^u  \ ,
\ee
where the four amplitudes are given by
\bea
M^c & =& ie_0g_{\Theta NK} \bar{\Theta}\gamma_\mu\gamma_5 A^\mu N K ,\nonumber\\
M^t & =&  \frac{ie_0g_{\Theta NK}}{t-M_K^2}\bar{\Theta}
\gamma_\mu\gamma_5 (q-k)^\mu (2q-k)_\nu A^\nu N K,\nonumber\\
M^s & =&  -g_{\Theta NK}\bar{\Theta} \gamma_\mu\gamma_5\partial^\mu K
\frac{[\gamma\cdot(k + P_i) +M_n]}{s-M_n^2} 
\left[ e_n\gamma_\alpha + \frac{i{\kappa}_n }{2M_n}\sigma_{\alpha\beta} 
k^\beta\right] A^\alpha N, \nonumber\\
M^u & =&  -g_{\Theta NK}\bar{\Theta} 
\left[ e_\theta\gamma_\alpha + \frac{i{\kappa}_\theta }{2M_\Theta}
\sigma_{\alpha\beta} k^\beta\right] A^\alpha 
\frac{[\gamma\cdot(P_f-k)+M_\Theta]}{u-M_\Theta^2} 
\gamma_\mu\gamma_5\partial^\mu K N,
\eea
where $e_0$ is the positive unit charge. 
The symbols, $N$ and $A$ denote the initial neutron and photon fields, while
$\bar{\Theta}$ and $K$ denote the final state $\Theta^+$ and kaon.
In the {\it s}-channel
the vector coupling vanishes since $e_n=0$. We define
the coupling constant $g_{\Theta NK}\equiv g_A M_n/f_\theta$ with 
the axial vector coupling $g_A=5/3$, while the decay constant is given by: 
\be
f_\theta=g_A\left(1-\frac{p_0}{E_n+M_n}\right) 
\left[ \frac{|{\bf p}^\prime|^3(E_n+M_n)}
{4\pi M_\Theta\Gamma_{\Theta^+\to K^+ n}}\right]^{1/2},
\ee
where ${\bf p}^\prime$ and $p_0$ are momentum and energy of the kaon
in the $\Theta^+$ rest frame, and $E_n$ is the energy of the neutron.
We adopt $g_{\Theta NK}=2.96$ corresponding to
$\Gamma_{\Theta^+\to K^+ n}=10$ MeV in the calculations.
The $\Theta^+$'s magnetic moment 
 $\mu_\theta=0.13(e_0/2M_\Theta)$ 
is estimated in Ref.~\cite{zhao-theta} based on the 
the diquark model of Jaffe and Wilczek~\cite{JW}, which is consistent 
with the detailed model studies of Refs.~\cite{h-d-c-z,l-h-d-c-z} 
and an estimate in Ref.~\cite{nam}.

The $K^*$ exchange is also considered in this work. 
The effective Lagrangian for $K^*K\gamma$ is given by 
\be
{\cal L}_{K^*K\gamma}=\frac{ie_0g_{K^*K\gamma}}{M_K}
\epsilon_{\alpha\beta\gamma\delta}
\partial^\alpha A^\beta\partial^\gamma V^\delta K  + \mbox{H.c.} \ ,
\ee
where $V^\delta$ denotes the $K^*$ field; $g_{K^*K\gamma}=0.744$ is  
determined by the $K^{*\pm}$ decay width 
$\Gamma_{K^{*\pm}\to K^\pm\gamma}=50$ keV~\cite{pdg2000}.

The $K^*N\Theta$ interaction is given by 
\be 
{\cal L}_{\Theta N K^*}=g_{\Theta N K^*}\bar{\Theta}
(\gamma_\mu +\frac{\kappa_\theta^*}{2M_\Theta}
\sigma_{\mu\nu}\partial^\nu)V^\mu N   + \mbox{H.c.} \ ,
\ee
where $g_{\Theta N K^*}$ and $\kappa_\theta^*$ denote
the vector and tensor couplings, respectively.
So far, there is no experimental information about these two 
couplings. A reasonable assumption based on 
an analogue to vector meson exchange in pseudoscalar meson production
is that $|g_{\Theta N K^*}|=|g_{\Theta N K}|$.
For the tensor coupling, we assume $|\kappa_\theta^*|=|\kappa_\rho|= 3.71$, 
the same as $\rho NN$ tensor coupling but with an arbitary phase.
Therefore, four sets of different phases are possible.

%%%%-- Parity negative

\begin{figure}
\begin{center}
\epsfig{file=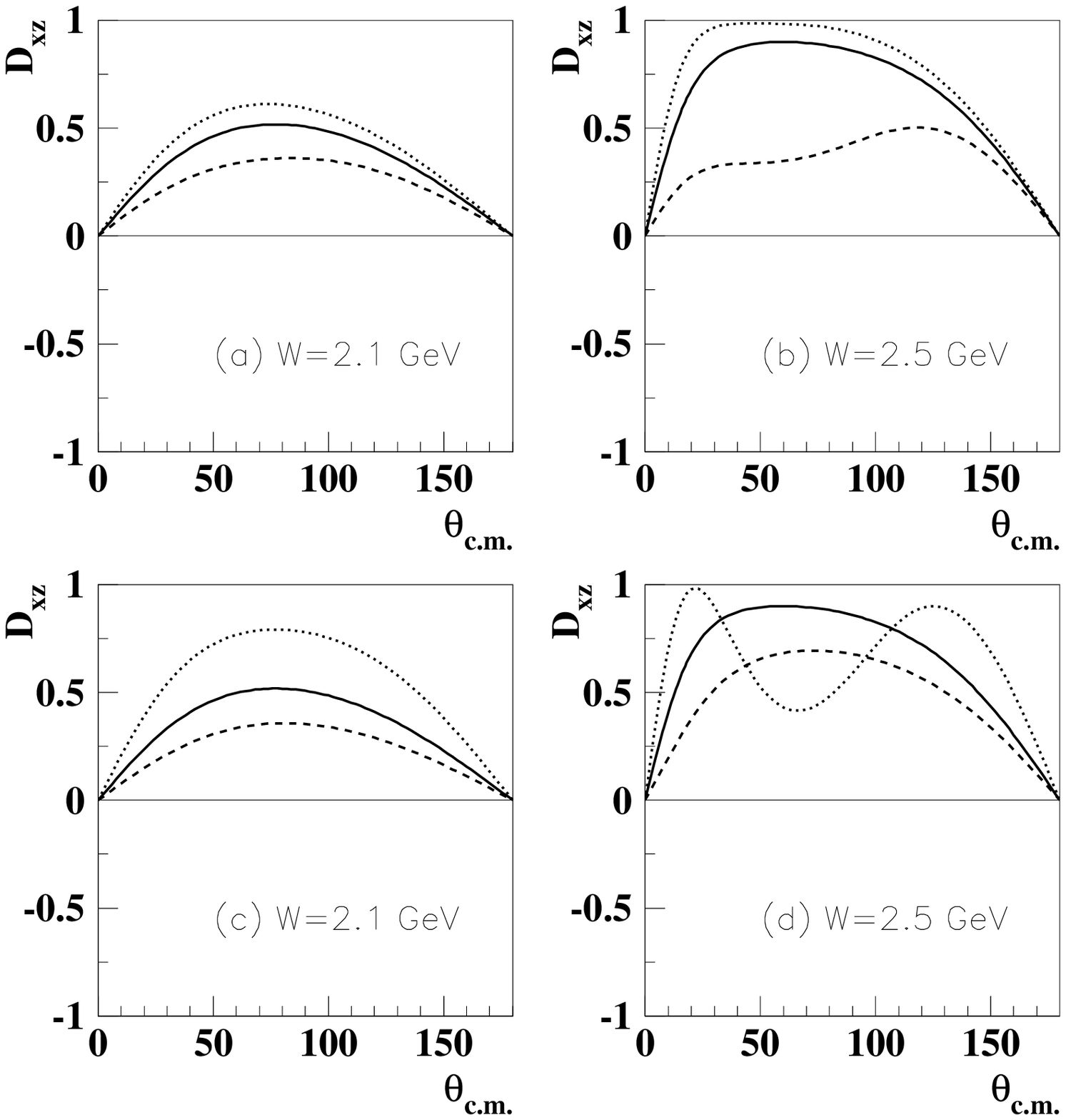, width=10cm,height=10.cm}
\caption{Beam-target double polarization asymmetry for 
$\Theta^+$ of $1/2^+$ at $W=2.1$ and 2.5 GeV. 
The solid curves are results in the Born limit, while the dashed and
dotted curves denote results with the $K^*$ exchange included 
with different phases: $(g_{\Theta N K^*}, \kappa_\theta^*)= (-2.8,-3.71)$ 
(dashed curves in (a) and (b)),
$(+2.8,+3.71)$ (dotted curves in (a) and (b)), 
$(-2.8, +3.71)$ (dashed curves in (c) and (d)), and 
$(+2.8, -3.71)$ (dotted curves in (c) and (d)).}
\protect\label{fig:(2)}
\end{center}
\end{figure}

For $\Theta^+$ of $1/2^-$, the effective Lagrangian for the $\Theta^+ n K$
system conserves parity and is gauge invariant. 
This suggests that the electromagnetic interaction will not contribute
to the contact term, which can be also seen in the leading 
term of the nonrelativistic expansion, where the derivative operator
is absent. The Born approximation therefore includes three transitions
(Fig.~\ref{fig:(2)}b, c and d):
The invariant amplitude can be written as
\be
{\cal M}_{fi}=M^t + M^s + M^u  \ ,
\ee
where the three transitions are given by
\bea
M^t & =&  -\frac{e_0g_{\Theta NK}}{t-M_K^2}\bar{\Theta}
(2q-k)_\mu A^\mu N, \nonumber\\ 
M^s & =&  g_{\Theta NK}\bar{\Theta} K
\frac{[\gamma\cdot(k + P_i) +M_n]}{s-M_n^2} 
\left[ e_n\gamma_\mu + \frac{i{\kappa}_n }{2M_n}\sigma_{\mu\nu} k^\nu\right]
A^\mu N,
\nonumber\\
M^u & =&  g_{\Theta NK}\bar{\Theta} 
\left[ e_\theta\gamma_\mu + \frac{i{\kappa}_\theta }{2M_\Theta}
\sigma_{\mu\nu} k^\nu\right] A^\mu 
\frac{[\gamma\cdot(P_f-k)+M_\Theta]}{u-M_\Theta^2} 
 N K, 
\eea
where the coupling constant 
$g_{\Theta NK}=[4\pi M_\Theta \Gamma_{\Theta^+\to K^+ n}
/|{\bf p}^\prime|(E_n+M_n)]^{1/2}$ 
has a value of 0.61 
corresponding to $\Gamma_{\Theta^+\to K^+ n}=10$ MeV considered in this work.

If $\Theta^+$ has spin-parity $1/2^-$, one may simply estimate its
magnetic moment as the sum of $(u\bar{s})$ and $(udd)$ clusters, 
since the relative orbital angular momentum is zero. 
Assuming these two clusters are both color singlet, the 
total magnetic moment of this system can be written as
\be
\mu_\theta =\left(\frac{2e_0}{6m_u}+\frac{e_0}{6m_s}\right)
-\frac{e_0}{3m_u}=\frac{e_0}{6m_s},
\ee
which leads to a small anomalous magnetic moment 
(since $3m_s\simeq M_\Theta$).
However, since such a simple picture may not be sufficient, we 
also include $\kappa_\theta=\pm\kappa_p=\pm 1.79$ to make a sensitivity
test.

We also include $K^*$ exchange in the $1/2^-$ production, in which
the $K^*N\Theta$ interaction is given by 
\be 
{\cal L}_{\Theta N K^*}=g_{\Theta N K^*}\bar{\Theta}
\gamma_5(\gamma_\mu +\frac{\kappa_\theta^*}{2M_\Theta}
\sigma_{\mu\nu}\partial^\nu)V^\mu N   + \mbox{H.c.} \ .
\ee
As in the $1/2^+$ case, we assume
$|g_{\Theta N K^*}|=|g_{\Theta N K}|=0.61$. 
In the calculation we choose, somewhat arbitrarily, the anomalous
magnetic moment $|\kappa_\theta^*|=0.37$, which is one order of
magnitude smaller than that in the production of $1/2^+$. Of course, we
could have chosen it to be in the same ratio as that of $g_{\Theta N
K}$ in the two cases (about a factor of 5).  However, the precise value
of $|\kappa_\theta^*|$ does not alter the overall conclusions of this
paper regarding the qualititative features of the BT asymmetry near
threshold.

%%%%%%

In the helicity frame, the transition amplitude 
can be written as:
\be
T_{\lambda_\theta,\lambda_\gamma \lambda_N}\equiv
\langle \Theta^+, \lambda_\theta, {\bf P}_\theta; K^-, \lambda_0, {\bf q} 
| \hat{T} | n, \lambda_N, {\bf P}_i; \gamma, \lambda_\gamma, {\bf k}\rangle ,
\ee
where $\lambda_\gamma=\pm 1$, $\lambda_N\pm 1/2$, 
$\lambda_0=0$, and $\lambda_\theta$ are helicities 
of photon, neutron, $K^-$, and $\Theta^+$, respectively.
Following the convention of 
Ref.~\cite{walker}, we define the four independent helicity as:
\bea
H_1 &=& 
T_{-\frac 12, 1 -\frac 12}, \nonumber\\
H_2 &=& 
T_{-\frac 12, 1 \frac 12}, \nonumber\\
H_3 &=&
T_{\frac 12, 1 -\frac 12}, \nonumber\\
H_4 &=&
T_{\frac 12, 1 \frac 12}.
\eea

The BT polarization asymmetry is defined as the ratio 
between the polarized cross section and unpolarized one
in terms of the total c.m. scattering angle $\theta_{c.m.}$ (the angle
between ${\bf k}$ and ${\bf q}$).
Analytically, it can be related to the 
density matrix elements for $\Theta^+\to K^+ n$. 
We are interested in 
the observable that has the photon circularly polarized along 
the photon moment direction $\hat{z}$ and the neutron transversely 
polarized along the $\hat{y}$-axis perpendicular to the reaction plane.
The $\Theta^+$ decay density matrix element
is defined as
\be
\rho^{BT}_{\lambda_\theta\lambda_\theta^\prime}
=\frac{1}{2N}\sum_{\lambda_\gamma, \lambda_N}
\lambda_\gamma 
T_{\lambda_\theta,\lambda_\gamma -\lambda_N} 
T^*_{\lambda_\theta^\prime,\lambda_\gamma \lambda_N}  ,
\ee
where, $N=\frac 12 \sum_{\lambda_\theta,\lambda_\gamma,\lambda_N}
|T_{\lambda_\theta,\lambda_\gamma \lambda_N}|^2$ is the normalization
factor. Because of parity conservation and the requirement of
Hermiticity, only two elements are independent: 
\bea
\rho^{BT}_{\frac 12,\frac 12} &=& \rho^{BT}_{-\frac 12,-\frac 12},\nonumber\\
\rho^{BT}_{-\frac 12,\frac 12} &=& -\rho^{BT}_{\frac 12,-\frac 12},
  \ \ \ \mbox{with} \ \ \ 
\mbox{Re}\rho^{BT}_{-\frac 12,\frac 12}
=\mbox{Re}\rho^{BT}_{\frac 12,-\frac 12}=0 \ .
\eea
This leads to an expression for the BT asymmetry,
\be
\label{bt-heli}
{\cal D}_{xz}=\frac{\rho^{BT}_{\frac 12,\frac 12}}{\rho^0_{\frac 12,\frac 12}}
=\frac{1}{N}\mbox{Re}\{H_1 H^*_2 +H_3 H^*_4\} ,
\ee
where $\rho^0_{\frac 12,\frac 12}$ is the unpolarized density matrix
element, and the subscript $xz$ denotes the polarization direction of
the the initial neutron target along $x$-axis in the production plane
and the incident photon along the $z$-axis.

The expression of BT asymmetry for $1/2^+$ and $1/2^-$ 
is the same. But the underlying dynamics will be determined
by the parities, and leads to different behaviors of ${\cal D}_{xz}$.
For the production of $1/2^+$ and $1/2^-$ respectively, we will show that
analytical features due to the dynamics
arise from the BT asymmetry, and turn out to be 
quite different for the two parity cases.

For $1/2^+$, the transition amplitudes can be 
expressed in terms of the CGLN amplitudes~\cite{CGLN}:
\be
\langle \Theta^+, \lambda_\theta, {\bf P}_\theta; K^-, \lambda_0, {\bf q} 
| \hat{T} | n, \lambda_N, {\bf P}_i; \gamma, \lambda_\gamma, {\bf k}\rangle 
=\langle \lambda_\theta | {\bf J}\cdot \veps_\gamma |\lambda_N \rangle  \ ,
\ee
where the operator ${\bf J}\cdot \veps_\gamma$ has a form:
\be
\label{heli-posi}
{\bf J}\cdot \veps_\gamma= if_1\vsig \cdot \veps_\gamma
+ f_2\frac{1}{|{\bf q}||{\bf k}|}\vsig\cdot{\bf q}
\vsig\cdot({\bf k}\times\veps_\gamma)
+if_3\frac{1}{|{\bf q}||{\bf k}|}\vsig\cdot{\bf k}{\bf q}\cdot \veps_\gamma
+if_4\frac{1}{|{\bf q}|^2}\vsig\cdot{\bf q}{\bf q}\cdot \veps_\gamma .
\ee
The coefficients $f_{1,2,3,4}$ are functions of energies, momenta, 
and scattering angle $\theta_{c.m.}$, and contain information on dynamics.
They provide an alternative expression for the BT asymmetry:
\be
\label{bt-posi}
{\cal D}_{xz}=\sin\theta_{c.m.}\mbox{Re}\{
f_1 f_3^* - f_2 f_4^* 
+ \cos\theta_{c.m.} (f_1 f_4^* -f_2 f_3^*)\} .
\ee
Note that the above 
expression is the same as Eq. (B8) of Ref.~\cite{fasano92}.

Near threshold, useful analytical information can be obtained. 
It is well-established that the contact and {\it t}-channel terms
are the leading contributions. In particular, the 
Kroll-Ruderman term ($f_1$), is dominant in the transition. 
Compared to $f_3$, amplitude $f_4$ is relatively suppressed 
by a further power on 
the (small) final-state three momentum $|{\bf q}|$.
Amplitude $f_2$ is also relatively suppressed in comparison with $f_3$
as it comes from the magnetic transition. Therefore, 
$f_1 f_4^*$ and $f_2 f_3^*$ will both be relatively suppressed 
in comparison with $f_1 f_3^*$ and further suppressed by $\cos\theta_{c.m.}$
in the middle angles. 
Thus, the behaviour of ${\cal D}_{xz}$ near threshold can be 
approximated by 
\be 
{\cal D}_{xz}\simeq\sin\theta_{c.m.}
\mbox{Re}\{ f_1 f_3^*\} .
\ee
Since the CGLN amplitudes only depend weakly on the scattering
angle $\theta_{c.m.}$ (via the Mandelstam variables), this approximation 
implies a $\sin\theta_{c.m.}$ behavior 
of ${\cal D}_{xz}$, and the sign of ${\cal D}_{xz}$ is determined by 
the product. 

In Fig.~\ref{fig:(2)}, as shown by the curves from full calculations 
at $W=2.1$ GeV, 
a clear $\sin\theta_{c.m.}$ behaviour appears in the BT asymmetry.
Although the sign change of the  $K^*$ exchange results in 
a quite significant change to the asymmetry values, the curves
nevertheless confirm the dominance of the $f_1 f_3^*$ term 
in Eq.~(\ref{bt-posi}). 
This result is by no means trivial. It suggests that 
although the $K^*$ exchange might produce significantly different 
predictions for the cross sections~\cite{zhao-theta}, 
the BT asymmetry will tend to 
behave predominantly as $\sin\theta_{c.m.}$
with its sign determined by $f_1 f_3^*$.
Such a feature can be regarded as a signature 
of $1/2^+$ spin-parity for the $\Theta^+$.

It is natural to expect that 
such a behavior should not hold at higher energies, where 
other mechanisms may contribute, and $f_1$ and $f_3$ will no longer 
be the leading terms. As shown by the asymmetries for different 
$K^*$ exchange phases at $W=2.5$ GeV, structures deviating from 
$\sin\theta_{c.m.}$ arise. In particular, comparing the dashed curve in 
Fig.~\ref{fig:(2)} (b) and the dotted one in Fig.~\ref{fig:(2)}(d)
with the solid curve for the exclusive Born terms, the structures
reflect the strong interference from the $K^*$ exchange
via the $\cos\theta_{c.m.}(f_1 f_4^*-f_2 f_3^*)$ term in Eq.~(\ref{bt-posi}).

Similar analysis can be applied to the production of $1/2^-$. 
In general, the transition amplitude can be arranged
in a way similar to the CGLN amplitude:
\be
{\bf J}\cdot\veps_\gamma=
iC_1\frac{1}{|{\bf k}|}\vsig\cdot(\veps_\gamma\times{\bf k})
+C_2\frac{1}{|{\bf q}|}\vsig\cdot{\bf q}\vsig\cdot\veps_\gamma
+iC_3\frac{1}{|{\bf q}||{\bf k}|^2}\vsig\cdot{\bf k}
{\bf q}\cdot (\veps_\gamma\times{\bf k})
+iC_4\frac{1}{|{\bf q}|^2|{\bf k}|}\vsig\cdot{\bf q}
{\bf q}\cdot (\veps_\gamma\times{\bf k}) ,
\ee
where coefficients $C_{1,2,3,4}$ are functions of energies, momenta and 
scattering angle, and contain dynamical information on the transitions. 
Restricted to the kinematics near threshold, 
a term proportional to 
${\bf q}\cdot\veps_\gamma\vsig\cdot({\bf k}\times{\bf q})$ 
in the {\it u}-channel is neglected since it comes from a  
higher order contribution.
An advantage of this formulation is that one can 
express the above in parallel to the CGLN amplitudes.
Since  $(\veps_\gamma\times\hat{\bf k})= i\lambda_\gamma\veps_\gamma$, 
one can replace vector $(\veps_\gamma\times\hat{\bf k})$
with $ i\lambda_\gamma\veps_\gamma$, and rewite the operator as
\be
{\bf J}\cdot\veps_\gamma=i\lambda_\gamma\left[iC_1\vsig \cdot \veps_\gamma
+ C_2\frac{1}{|{\bf q}||{\bf k}|}\vsig\cdot{\bf q}
\vsig\cdot({\bf k}\times\veps_\gamma)
+iC_3\frac{1}{|{\bf q}||{\bf k}|}\vsig\cdot{\bf k}{\bf q}\cdot \veps_\gamma
+iC_4\frac{1}{|{\bf q}|^2}\vsig\cdot{\bf q}{\bf q}\cdot \veps_\gamma \right],
\ee
which has exactly the same form as Eq.~\ref{heli-posi} apart from an overall
phase factor from the photon polarization $i\lambda_\gamma$.
It also suggests that the BT asymmetry for $1/2^-$ in terms of those 
coefficients has the same form as Eq.~\ref{bt-posi}:
\be
{\cal D}_{xz}=\sin\theta_{c.m.}
\mbox{Re}\{C_1 C_3^* -C_2 C_4^* +\cos\theta_{c.m.}
(C_1 C_4^*-C_2 C_3^*) \} ,
\ee
which is indeed the case. Quite remarkably, 
the behaviour of ${\cal D}_{xz}$ due to these two different parities 
now becomes more transparent since the role played by 
the dynamics has been isolated out.

Note that since ${\cal D}_{xz}$ is roughly proportional to $\sin\theta_{c.m.}$, 
it will vanish at $\theta_{c.m.}=0^\circ$ and $180^\circ$, and 
any sign change will reflect the competition
among the $C$ coefficients due to the dynamics.

The most important difference between the $1/2^-$ and $1/2^+$ cases is that 
the role  of $C_1$ may not be as significant as $f_1$ in the 
production of $1/2^+$. 
In the Born approximation limit, the term of $C_1$, though dominant,
comes from the {\it s}- and {\it u}-channel, which differs from 
the Kroll-Ruderman contribution from the contact term in $1/2^+$
production. 
As a result, a relative suppression from the baryon propagator 
is expected when the energy increases. 
One thus may conjecture that other mechanisms, such as $K^*$ exchange, 
may easily compete with the Born contribution
and produce sign changes to the asymmetries above threshold. 

\begin{figure}
\begin{center}
\epsfig{file=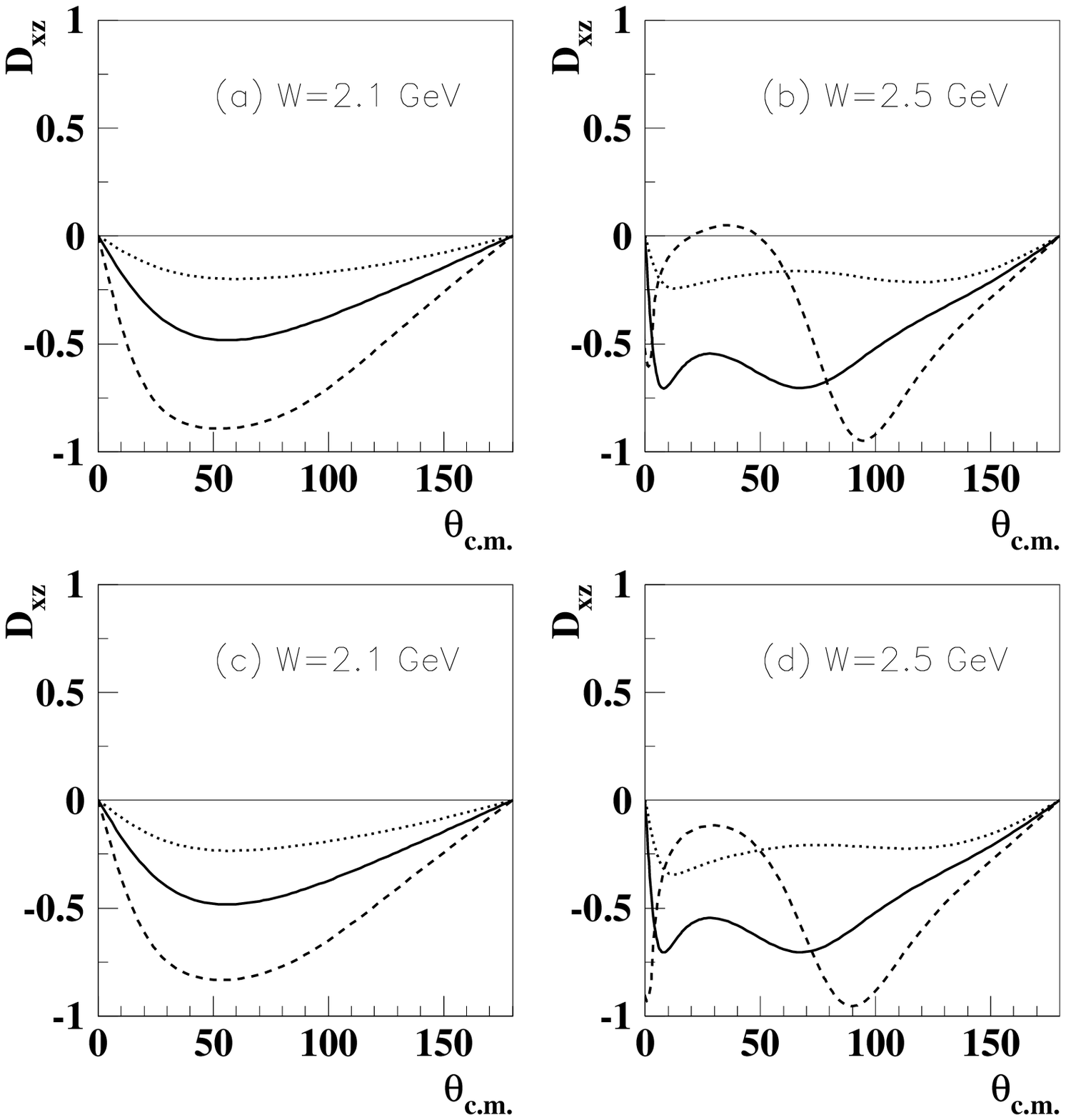, width=10cm,height=10.cm}
\caption{Beam-target double polarization asymmetry for 
$\Theta^+$ of $1/2^-$ at $W=2.1$ and 2.5 GeV. 
The solid curves are results in the Born limit, while the dashed and
dotted curves denote results with the $K^*$ exchange included 
with different phases: $(g_{\Theta N K^*}, \kappa_\theta^*)= (-0.61,-0.371)$ 
(dashed curves in (a) and (b)),
$(+0.61,+0.371)$ (dotted curves in (a) and (b)), 
$(-0.61, +0.371)$ (dashed curves in (c) and (d)), and 
$(+0.61, -0.371)$ (dotted curves in (c) and (d)).}
\protect\label{fig:(3)}
\end{center}
\end{figure}

However, as shown by the solid curve at $W=2.1$ GeV
in Fig.~\ref{fig:(3)}, 
the BT asymmetry in the Born limit exhibits a $\sin\theta_{c.m.}$
behavior, which suggests the dominance
of either $C_1 C_3^*$ or $C_2 C_4^*$ near threshold, and
the $\mbox{Re}\{\cos\theta_{c.m.}(C_1 C_4^*-C_2 C_3^*)\}$
term should not be important.
The inclusion of the $K^*$ exchange certainly does not change
this situation as shown by the dashed and dotted curves at $W=2.1$ GeV.
In particular, an absolute sign difference,  unlikely to be accidental,
appears between these two parities, and needs to be understood.

First, let us try to understand the behavior 
of ${\cal D}_{xz}$ in $1/2^+$ production.
As mentioned previously, the BT asymmetry for $1/2^+$
is controlled by $f_1 f_3^*$ in the Born terms. 
A detailed analysis gives the leading term of this combination:
\be
\label{approx-1}
f_1 f_3^* \simeq 
-e_0^2g_{\Theta NK}^2
\frac{2}{u-M_\Theta^2}
{\cal F}_c(k, q){\cal F}_u(k, q) \ ,
\ee
where $f_1$ is dominated by the Kroll-Ruderman term 
and $f_3$ by the electric coupling in the {\it u}-channel. 
It is worth noting that the {\it s}-channel will not contribute
to $f_3$, and the magnetic coupling in the {\it u}-channel
is relatively suppressed. These features fix the sign of 
${\cal D}_{xz}$, and
underline the dominance of $f_1 f_3^*$ a dynamical consequence. 
${\cal F}_c(k, q)$ and ${\cal F}_u(k, q)$ are form factors
for the contact and {\it u}-channel, and are treated the same 
in this approach. This assumption may bring in uncertainties
at high energies, but should be a reasonable one near threshold.

For the production of $1/2^-$, we found that 
the $C_1 C_3^*$ term plays a dominant role in the asymmetry.
A detailed analysis shows that the dominant 
contribution to $C_1$ is from the {\it s}-channel, while the contribution
from the
{\it u}-channel is strongly suppressed by the limited kinematics.
A very important property arising from the  $C_3$ term
is that it will be dominated by the {\it t}-channel kaon exchange
via the decomposition ${\bf q}\cdot\veps_\gamma=
\vsig\cdot{\bf q}\vsig\cdot\veps_\gamma 
+i\vsig\cdot(\veps_\gamma\times\hat{\bf k}){\bf q}\cdot\hat{\bf k}
-i\vsig\cdot\hat{\bf k}{\bf q}\cdot(\veps_\gamma\times\hat{\bf k})$.
Therefore, we have
\be
C_1 C_3^* \simeq
-e_0^2g_{\Theta NK}^2 \frac{\kappa_n}{2M_n}\frac{2|{\bf k}||{\bf q}|}
{(W-M_n)(t-M_K^2)} {\cal F}_t(k, q) {\cal F}_s(k, q) ,
\ee
which will be  negative since $\kappa_n=-1.91$. 
In comparison with Eq.~\ref{approx-1},
it gives a dynamical reason for the sign difference 
between the two parities. Also, note that 
${\cal F}_t(k, q)$ and ${\cal F}_s(k, q)$ are form factors for 
the {\it t}- and {\it s}-channels, 
which is a further indication of the very different characteristic 
dynamics of these two parity cases probed by the double polarization asymmetry.

The dominance of $C_1 C_3^*$ near threshold implies 
that the BT asymmetry ${\cal D}_{xz}$ is not sensitive to the 
magnetic moment of the $\Theta^+$, since the {\it u}-channel 
contribution from the anomalous magnetic moment 
of $\Theta^+$ to $C_3$ is much smaller than the {\it t}-channel. 
We indeed see this by considering $\kappa_\theta=\pm 1.91$, 
which does not change the sign or basic features of ${\cal D}_{xz}$.

The dominance of $C_1 C_3^*$ near threshold
also holds even when $K^*$ exchange included. 
As shown by the dashed and dotted curves in Fig.~\ref{fig:(3)} at
$W=2.1$ GeV, $K^*$ exchange introduces significant interference
into the asymmetries, which does not, however, change the 
dominant $\sin\theta_{c.m.}$ behaviour. 
The explicit different signs of the BT asymmetry in the two cases, 
clearly distinguishes $1/2^-$ from $1/2^+$
parities near threshold.

Such a feature is undoubtedly remarkable, and experimentally sensible.
Recall that a strong $K^*$ coupling could lead to a significant enhancement
of the cross sections~\cite{hyodo,nam,liu-ko,oh,zhao-theta} 
for the production of $1/2^-$.
As a result, a single measurement of the cross section 
may not be sufficient for the determination of the 
$\Theta^+$'s parity. 
It turns out that the BT polarization asymmetry ${\cal D}_{xz}$
is able to picked out the most distinct dynamical differences
between $1/2^+$ and $1/2^-$, and distinguish these two parities 
in a measurement near threshold.

It is worth noting that a similar behaviour of the Born terms also 
exists  
in the polarized beam asymmetry between these two parities~\cite{zhao-theta}.
However, the differences picked out by this single polarization 
asymmetry turn out not to be as marked as 
in the double polarization. Near threshold, although
the Born terms will lead to differently signed asymmetries 
for $1/2^+$ and $1/2^-$~\cite{zhao-theta}, they 
will destructively interfere with
other mechanisms, such as the $K^*$ exchange, 
and lose their characteristic feature. 

In summary, we have analyzed the double polarization asymmetry,
${\cal D}_{xz}$, in $\gamma n\to \Theta^+ K^-$, and showed it
to be a useful filter for determining the parity
of $\Theta^+$, provided its spin-parity is either $1/2^+$ or $1/2^-$.
Due to dynamical reasons, asymmetry ${\cal D}_{xz}$
near threshold would exhibit a similar behaviour but opposite 
sign.
The advantage of studying polarization observables is that 
uncertainties arising from the unknown form factors 
can be partially avoided in a phenomenology. 
Therefore, although better knowledge of the form factors
will improve the quantitative predictions, 
it should not change the threshold behaviour of ${\cal D}_{xz}$
dramatically. 
However, special caution should be given to the roles played 
by a possible spin-$3/2$ partner in the {\it u}-channel, as well as 
{\it s}-channel nucleon resonances. In particular, 
as studied by Dudek and Close~\cite{d-c}, the spin-3/2 partner 
may have a mass close to the $\Theta^+$. Thus, a 
significant contribution from the spin-3/2 pentaquark state
may be possible. Its impact on the BT asymmetry needs to be investigated.
In brief, due to the lack of knowledge in this area, 
any results for the BT asymmetry would be extremely important 
for progress in gaining insights into the nature of pentaquark states
and dynamics for their productions. Experimental facilities 
at Spring-8, JLab, ELSA, and ESRF should have access to the BT asymmetry
observable.

The authors thank Frank E. Close for many useful discussions.
This work is supported by grants from 
the U.K. Engineering and Physical
Sciences Research Council (Grant No. GR/R78633/01).

\begin{thebibliography}{99}
%
\bibitem{spring-8} T. Nakano {\it et al.} [LEPS Collaboration],
	Phys. Rev. Lett. {\bf 91}, 012002 (2003).
%
\bibitem{diana} V. Barmin {\it et al.} [DIANA Collaboration],
	hep-ex/0304040.
%
\bibitem{clas} S. Stepanyan {\it et al.} [CLAS Collaboration],
	hep-ex/0307018.
%
\bibitem{saphir} J. Barth {\it et al.} [SAPHIR Collaboration],
	hep-ex/0307083.
%
\bibitem{clas-2} V. Kubarovsky {\it et al.} [CLAS Collaboration],
	hep-ex/0311046.
%
\bibitem{na49} C. Alt {\it et al.}, [NA49 Collaboration], 
	hep-ex/0310014.
%
\bibitem{skyrme} T.H.R. Skyrme, 
	Proc. Royal Soc. {\bf A260}, 127 (1961);
	Nucl. Phys. {\bf 31}, 556 (1962).
%
\bibitem{dpp} D. Diakonov, V. Petrov, and M. Polyakov, 
	Z. Phys. {\bf A 359}, 309 (1997).

%%%% Skyrme model
%
\bibitem{bfk} D. Borisyuk, M. Faber, and A. Kobushkin,
	hep-ph/0307370.
%
\bibitem{prasza-1} H.-C. Kim and M. Praszalowicz,
	hep-ph/0308242.
%
\bibitem{praszalowicz} M. Praszalowicz, 
	Phys.\ Lett.\ B {\bf 575}, 234 (2003); hep-ph/0308114;
	hep-ph/0311230.
%
\bibitem{cohen} T.D. Cohen and R.F. Lebed, 
	hep-ph/0309150.
%
\bibitem{ikor} N. Itzhaki, I.R. Klebanov, P. Ouyang, and L. Rastelli,
	hep-ph/0309305.
%
\bibitem{cohen-2} T.D. Cohen, hep-ph/0312191.
%
\bibitem{wu-ma} B. Wu and B.-Q. Ma, 
	hep-ph/0311331; hep-ph/0312041.
%
%%%% Quark model
%
\bibitem{gao-ma} H. Gao and B.Q. Ma,
	Mod. Phys. Lett. {\bf A 14}, 2313 (1999).
%
\bibitem{JW} R. Jaffe and F. Wilczek, 
        Phys. Rev. Lett. {\bf 91}, 232003 (2003); hep-ph/0307341.
%
%
\bibitem{k-l} M. Karliner and H.J. Lipkin, 
	hep-ph/0307343; hep-ph/0307243.
%
\bibitem{s-r} Fl. Stancu and D.O. Riska,
	Phys.\ Lett.\ B {\bf 575}, 242 (2003); hep-ph/0307010.
%
\bibitem{caps} S. Capstick, P.R. Page, and W. Roberts,
	Phys.\ Lett.\ B {\bf 570}, 185 (2003); hep-ph/0307019.
%
\bibitem{carlson} C.E. Carlson, C.D. Carone, H.J. Kwee, and V. Nazaryan,
	Phys.\ Lett.\ B {\bf 573}, 101 (2003); hep-ph/0307396.
%
\bibitem{cheung} K. Cheung, 
	hep-ph/0308176.
%
\bibitem{glozman} L. Ya. Glozman, 
	Phys.\ Lett.\ B {\bf 575}, 18 (2003); hep-ph/0308232.
%
\bibitem{jm} B.K. Jennings and K. Maltman, 
	hep-ph/0308286.
%
\bibitem{hzyz} F. Huang, Z.Y. Zhang, Y.W. Yu, and B.S. Zou,
	hep-ph/0310040.
%
\bibitem{okl} Y. Oh, H. Kim, and S.H. Lee, 
	hep-ph/0310117.
%
\bibitem{dp} D. Diakonov and V. Petrov,
	hep-ph/0310212.
%
\bibitem{bijker} R. Bijker, M.M. Giannini, and E. Santopinto,
	hep-ph/0310281.
%
\bibitem{l-h-d-c-z} Y.-R. Liu, P.-Z. Huang, W.-Z. Deng, X.-L. Chen, and S.-L. Zhu,
	hep-ph/0312074.
%
%%%% Others
%
\bibitem{kk} D.E. Kahana and S.H. Kahana, 
	hep-ph/0310026.
%
\bibitem{gk} S.M. Gerasyuta and V.I. Kochkin,
	hep-ph/0310225; hep-ph/0310227.
%
%%%% QCD sum rules
%
\bibitem{zhu} S.L. Zhu, Phys. Rev. Lett. {\bf 91}, 232002 (2003);
	hep-ph/0307345.
%
\bibitem{mnnrl} R.D. Matheus {\it et al.}, 
	hep-ph/0309001.
%
\bibitem{sdo} J. Sugiyama, T. Doi, and M. Oka,
	hep-ph/0309271.
%
\bibitem{h-d-c-z} P.-Z.~Huang, W.-Z.~Deng, X.-L.~Chen and S.-L.~Zhu,
	hep-ph/0311108.

%
%%%% Lattice QCD
%
\bibitem{sasaki} S. Sasaki,
	hep-lat/0310014.
%
\bibitem{cfkk} F. Csikor, Z. Fodor, S.D. Katz, and T.G. Kovacs, 
	hep-lat/0309090.
%%%

%
\bibitem{kishimoto} T. Kishimoto and T. Sato,
	hep-ex/0312003.
%
\bibitem{juengst} H.G. Juengst, 
	nucl-ex/0312019.

%%%%
%
\bibitem{close1} F.E. Close, 
	Talk given at Hadron2003; hep-ph/0311087.
%
\bibitem{d-c} J.J. Dudek and F.E. Close, hep-ph/0311258.
%
%%%% Reaction theory
%
\bibitem{pr} M.V. Polyakov and A. Rathke, 
        hep-ph/0303138.
%
\bibitem{hyodo} T. Hyodo, A. Hosaka, and E. Oset,
        nucl-th/0307105.
%
\bibitem{nam} S.I. Nam, A. Hosaka, and H.-Ch. Kim, 
        hep-ph/0308313.
%
\bibitem{liu-ko} W. Liu and C.M. Ko, 
        Phys.\ Rev.\ C {\bf 68}, 045203 (2003); 
        nucl-th/0308034; nucl-ph/0309023.
%
\bibitem{oh} Y. Oh, H. Kim, and S.H. Lee,
        hep-ph/0310019.
%
\bibitem{yu} B.-G. Yu, T.-K. Choi, and C.-R. Ji,
	nucl-th/0312075.
%
%%%% narrow Theta+ 
%
\bibitem{asw} R.A. Arndt, I.I. Strakovsky, and R.L. Workman, 
	Phys.\ Rev.\ C {\bf 68}, 042201 (2003); 
	nucl-th/0308012; nucl-th/0311030.
%
\bibitem{aasw} Ya. I. Azimov, R.A. Arndt, I.I. Strakovsky, and R.L. Workman, 
	nucl-th/0307088.
%
\bibitem{nussinov} S. Nussinov,
	hep-ph/0307357.
%
\bibitem{gothe} R.W. Gothe and S. Nussinov,
	hep-ph/0308230.
%
\bibitem{hk} J. Haidenbauer and  G. Krein,
	Phys.\ Rev.\ C {\bf 68}, 052201 (2003); hep-ph/0309243.
%
\bibitem{cckn} C.E. Carlson, C.D. Carone, H.J. Kwee, and V. Nazaryan,
	hep-ph/0312325.
%
%\bibitem{prasza} M. Praszalowicz,
%	hep-ph/0311230.


%%%% Polarization asymmetry
%
\bibitem{zhao-theta} Q. Zhao, 
	Phys. Rev. D in press; hep-ph/0310350.
%
\bibitem{nakayama} K. Nakayama and K. Tsushima, 
	hep-ph/0311112.
%
\bibitem{okl-2} Y. Oh, H. Kim, and S.H. Lee,
	hep-ph/0312229.
%
\bibitem{thh} A.W. Thomas, K. Hicks, and A. Hosaka,
	hep-ph/0312083.
%
\bibitem{hanhart} C. Hanhart {\it et al.}, 
	hep-ph/0312236.
%
%%%%
%
\bibitem{pdg2000} D.~E.~Groom {\it et al.}  (Particle Data Group),
%``Review of particle physics,''
Eur.\ Phys.\ J.\ C {\bf 15}, 1 (2000).
%
\bibitem{walker} R.L. Walker, 
	Phys. Rev. {\bf 182}, 1729 (1969).
%
\bibitem{CGLN} G.F. Chew, M.L. Goldberger, F.E. Low, and Y. Nambu,
        Phys. Rev. {\bf 106}, 1345 (1957).
%
\bibitem{fasano92} C.G. Fasano, F. Tabakin, and B. Saghai, 
        Phys. Rev. C {\bf 46}, 2430 (1992).

\end{thebibliography}
\end{document}